\documentclass[a4paper,11pt]{article}
\usepackage{pos}

\usepackage{csquotes}          
\usepackage{hyperref}          
\usepackage{booktabs}       

\newcommand{\be}{\begin{equation}}
\newcommand{\ee}{\end{equation}}
\newcommand{\bea}{\begin{eqnarray}}
\newcommand{\eea}{\end{eqnarray}}

\newcommand{\bra}{\langle}
\newcommand{\ket}{\rangle}

\newcommand{\half}{\frac{1}{2}}

\newcommand{\qqquad}{\qquad\quad}
\renewcommand{\Re}{\mbox{Re}\,}
\renewcommand{\Im}{\mbox{Im}\,}

\newcommand{\eps}{\epsilon}

\newcommand{\bean}{\begin{eqnarray*}}
\newcommand{\eean}{\end{eqnarray*}}

\newcommand{\bignorm}{\big|\big|}

\newcommand{\E}{\mathbb{E}}

\title{Diffusion models learn distributions generated by complex Langevin dynamics}

\ShortTitle{Diffusion models and complex Langevin dynamics}

\author*[a]{Diaa E.\ Habibi}
\author[a]{Gert Aarts}
\author[b]{Lingxiao Wang}
\author[c,d]{Kai Zhou}

\affiliation[a]{Department of Physics, Swansea University, Swansea, SA2 8PP, United Kingdom}
\affiliation[b]{Interdisciplinary Theoretical and Mathematical Sciences Program (iTHEMS), RIKEN, Wako, Saitama 351-0198, Japan}
\affiliation[c]{School of Science and Engineering, The Chinese University of Hong Kong, Shenzhen (CUHK-Shenzhen), Guangdong, 518172, China}
\affiliation[d]{Frankfurt Institute for Advanced Studies, Ruth Moufang Strasse 1, D-60438, Frankfurt am Main, Germany\\
\mbox{}
}

\emailAdd{n.e.habibi@swansea.ac.uk, g.aarts@swansea.ac.uk, lingxiao.wang@riken.jp,  zhoukai@cuhk.edu.cn}

\abstract{
The probability distribution effectively sampled by a complex Langevin process for theories with a sign problem is not known a priori and notoriously hard to understand. Diffusion models, a class of generative AI, can learn distributions from data. In this contribution, we explore the ability of diffusion models to learn the distributions created by a complex Langevin process.
}

\FullConference{The 41st International Symposium on Lattice Field Theory (Lattice 2024)\\
28 July - 3 August, 2024\\
Liverpool, UK \\
}

\begin{document}
\maketitle

\section{Introduction}

Theories with a complex Boltzmann weight are hard to simulate using conventional numerical methods based on importance sampling, due to the sign and overlap problems \cite{Troyer:2004ge}. A prime example is QCD at nonzero baryon density, in which the quark determinant is complex for real quark chemical potential \cite{deForcrand:2009zkb,Aarts:2015tyj},
\be
\left[\det M(\mu)\right]^* = \det M(-\mu^*) \in \mathbb C.
\ee
Complex Langevin (CL) dynamics, in which the degrees of freedom are analytically extended, provides a potential solution, as it does not rely on importance sampling but explores a complexified manifold via a stochastic process \cite{parisi,klauder}. It is an extension of stochastic quantisation \cite{Parisi:1980ys,Damgaard:1987rr}, which is equivalent to path integral quantisation. 
CL has been shown to work in lattice field theories in three \cite{Aarts:2011zn} and four \cite{Aarts:2008wh} Euclidean dimensions with a severe sign problem, including in QCD \cite{Seiler:2012wz,Sexty:2013ica,Aarts:2016qrv,Sexty:2019vqx,Scherzer:2020kiu}, but it may also fail, even in simple models 
\cite{Ambjorn:1985iw,Aarts:2010aq,Aarts:2012ft}.
This situation was clarified a few years ago \cite{Aarts:2009uq,Aarts:2011ax,Nishimura:2015pba}
by the derivation of the formal relation between the complex distribution on the real manifold and the real and positive distribution on the complexified manifold, which is effectively sampled during the CL process, leading to practical criteria for correctness which need to be verified a posteriori. Nevertheless, issues remain and the reliability of the method depends on a precise understanding of the behaviour of the distribution at infinity and near poles in the CL drift. Recent work can be found in e.g.\ Refs.~\cite{Scherzer:2018hid,Scherzer:2019lrh,Alvestad:2022abf,Seiler:2023kes,Hansen:2024lkn}.

As should be clear, a crucial role is played by the distribution on the complexified manifold.
Unfortunately, this distribution turns out to be elusive, as the Fokker-Planck equation (FPE) linked to the CL process cannot be solved in general. In fact, even convergence is hard to understand, except in some simple cases, such as Gaussian models \cite{Aarts:2009hn} and models in which one can prove the dynamics takes place in a strip (see below) \cite{Aarts:2013uza}. A better characterisation of the distribution would therefore be welcome.

Diffusion models \cite{sohl-dickstein:2015deep,ho:2020denoising,SongErmon2020:generativemodels,Song:2021scorebased,yang:2022diffusion,song2021maximumlikelihoodtrainingscorebased,karras2022elucidatingdesignspacediffusionbased} 
are a class of generative AI, which learn distributions from data. They are widely popular and used in e.g.\ DALL-E \cite{2022arXiv220406125R} and Stable Diffusion \cite{Rombach_2022_CVPR}. The methodology of diffusion models relies on a stochastic process, similar to stochastic quantisation, but instead of using a known drift term derived from the underlying distribution, it learns the drift from data previously generated or collected. We have recently explored the relation between diffusion models and stochastic quantisation in scalar \cite{Wang:2023exq,Wang:2023sry} and U(1) gauge theories \cite{Zhu:2024kiu}, and studied the evolution of higher-order cumulants in detail \cite{Aarts:2024rsl}. Further connections between diffusion models and field theory are pointed out in Refs.~\cite{Hirono:2024zyg,Fukushima:2024oij}.

Given the success of diffusion models to learn distributions and the elusiveness of the distribution sampled in the CL process, it makes sense to combine these two approaches to deepen our understanding of the latter. In the next two sections we first remind the reader of CL dynamics for theories with a complex Boltzmann distribution and then briefly introduce diffusions models. In Sec.\ \ref{sec:examples} we then combine these in two simple cases with a single degree of freedom: the exactly solvable Gaussian case with a complex mass parameter, and the quartic model with a complex mass parameter, for which the distribution can be proven to be confined to a strip in the complex plane.

\section{Complex Langevin dynamics}

Consider one degree of freedom $x$, with a Boltzmann weight $\rho(x)$, such that
\be
\bra O(x)\ket = \int dx\, \rho(x)O(x), \qqquad \rho(x) = \frac{1}{Z} \exp[-S(x)], \qqquad Z = \int dx\,\rho(x).
\ee
The Langevin process and drift read
\be
\label{eq:L}
\dot x(t) = K[x(t)] +\eta(t), \qqquad K(x) = \frac{d}{dx} \log\rho(x) = -\frac{dS(x)}{dx},
\ee
where the dot indicates the Langevin time derivative and the noise satisfies $\bra \eta(t)\eta(t')\ket=2\delta(t-t')$. The corresponding FPE is
\be
\partial_t \rho(x;t) = \partial_x\left[ \partial_x -K(x) \right] \rho(x;t),
\ee
and for a real Boltzmann weight and drift this process converges to the stationary solution $\rho(x)$, typically exponentially fast \cite{Damgaard:1987rr}.

When the weight is complex, one may extend $x\to z=x+iy$ into the complex plane and write
\be
\dot z(t) = K[z(t)] +\eta(t), \qqquad K(z) = \frac{d}{dz} \log\rho(z) = -\frac{dS(z)}{dz}.
\ee
However, in this case the FPE cannot be used to show convergence as the corresponding Fokker-Planck Hamiltonian is no longer semi-positive definite \cite{Damgaard:1987rr}. 

We may consider the CL process,
\begin{align}
\dot x(t) = K_x +\eta_x(t), & \qqquad K_x = \Re \frac{d}{dz} \log\rho(z), \qquad \bra\eta_x(t)\eta_x(t')\ket=2N_x\delta(t-t'), \\
\dot y(t) = K_y+\eta_y(t),  & \qqquad K_y = \Im \frac{d}{dz} \log\rho(z), \qquad\bra\eta_y(t)\eta_y(t')\ket=2N_y\delta(t-t'),
\end{align}
with noise satisfying $N_x-N_y=1$.
The FPE equation for this process reads
\be
\label{eq:FPE}
\partial_t P(x,y;t) = \left[\partial_x\left(N_x\partial_x - K_x\right) + \partial_y\left(N_y\partial_y - K_y\right) \right] P(x,y;t),
\ee
such that
\be
\bra O[x(t)+iy(t)]\ket_\eta = \int dxdy\, P(x,y;t)O(x+iy).
\ee
It is preferable to consider real noise, $N_x=1, N_y=0$ \cite{Aarts:2009uq}.

The CL process yields the correct answer if a stationary solution to this FPE exists, such that 
\be
\int dxdy\, P(x,y)O(x+iy) = \int dx\, \rho(x) O(x),
\ee
or, shifting the integration variables at a formal level, 
\be
\label{eq:rhoP}
\rho(x) = \int dy\, P(x-iy,y).
\ee
Considerable effort has been invested in deriving criteria for correctness related to the behaviour of $P(x,y)$ at infinity and near poles of the drift (if there are any), which can be used a posteriori to justify the results \cite{Aarts:2009uq,Aarts:2011ax,Nishimura:2015pba,Scherzer:2018hid,Scherzer:2019lrh,Alvestad:2022abf,Seiler:2023kes,Hansen:2024lkn}. A better understanding of $P(x,y)$ in the stationary limit would therefore be very welcome. We emphasise that unlike the original weight $\rho(z)$, $P(x,y;t)$ is real and semi-positive definite, as it represents the real Langevin process in the two-dimensional plane.

\section{Diffusion models}

Diffusion models, as a class of probabilistic generative models, gradually corrupt data with incrementally increasing noise and are trained to reverse the process to build a generative model of the data
\cite{sohl-dickstein:2015deep,ho:2020denoising,SongErmon2020:generativemodels,Song:2021scorebased,yang:2022diffusion}.
We use the description in terms of stochastic differential equations (SDEs) \cite{Song:2021scorebased,song2021maximumlikelihoodtrainingscorebased,karras2022elucidatingdesignspacediffusionbased} and follow the notation of Ref.~\cite{Aarts:2024rsl}.

The diffusion process consists of two parts. The noise-injecting or forward process is described by the following SDE,
\be
\dot{x}(t) = K(x(t), t) + g(t)\eta(t),
\label{eq:forward_process}
\ee
where $K(x(t), t)$ is a drift term, $\eta\sim\mathcal{N}(0,1)$ is Gaussian noise and $g(t)$, the diffusion coefficient, is the time-dependent noise strength. The initial conditions for this process are determined by the target distribution $x(0) = x_0 \sim P_0(x_0)$ and the process runs between $0\leq t \leq T$. Properties of the distribution at the end of this process, $P(x,T)$, have been studied in Ref.~\cite{Aarts:2024rsl}. 

The second part corresponds to the denoising or backward process. Written in terms of reverse time $\tau = T - t$, the SDE reads \cite{anderson:1982reversetime}
\be
x'(\tau) = -K(x(\tau), T-\tau) + g^2(T-\tau)\partial_x \log P(x, T-\tau) + g(T-\tau)\eta(\tau),
\label{eq:backward_process}
\ee
with $0\leq\tau\leq T$. Initial conditions are sampled from a normal distribution with a variance comparable to the variance obtained at the end of the forward process. 
The so-called score, $\partial_x \log P(x,t)$, 
is not known a priori and is approximated by a quantity $s_\theta(x,t)$, which is determined, or `learnt', during the forward process via score matching \cite{Hyvarinen:2005jmlr}. 
Given a sample dataset of the target distribution, a score-based model can be trained starting from the Fisher divergence \cite{Song:2021scorebased},
 \be
 \mathcal{L}(\theta, \lambda) := \frac{1}{2}\int_0^T dt\, \E_{P(x,t)}\left[\lambda(t) \bignorm s_\theta(x,t) - \nabla\log P(x,t)\bignorm^2_2\right],
  \label{eq: weighted_training_objectice}
 \ee
where the weight $\lambda(t)$ is chosen to be the variance of the noise at time $t$.

After the diffusion model has been trained, new samples from the target distribution can be generated  by numerically solving the backward stochastic process (\ref{eq:backward_process}), substituting in the trained score model $s_\theta^*(x,t)$. Using a simple discretisation with stepsize $\Delta\tau$ one solves, for $0\leq \tau \leq T$,
\be
x_{\tau + \Delta \tau} = x_\tau +\left[ - K(x_\tau, T-\tau) + g^2(T-\tau)s_\theta^*(x_\tau, T-\tau) \right] \Delta\tau + g(T-\tau)\sqrt{\Delta\tau}\ \eta_\tau,
\ee
where $\eta_\tau \sim \mathcal{N}(0,1)$. This is indeed remarkably similar \cite{Wang:2023exq} to the formulation of stochastic quantisation, in which, however, the drift is time-independent and derived from a known distribution, as in Eq.~(\ref{eq:L}), rather than being learnt from data.

\section{Application to complex Langevin dynamics}
\label{sec:examples}

As stated above, for complex actions CL dynamics is capable of generating configurations but the corresponding probability distribution $P(x,y)$ is typically not available. This makes it hard to fully assess the reliability of the approach. A diffusion model, however, can learn (the log-derivative of) this distribution, in the form of the score $\nabla\log P(x,y)$. Notably, for the diffusion model it is irrelevant what the origin of the configurations is. 
The learned score can subsequently be used to study aspects of convergence of the CL process or to generate additional configurations.

To investigate the viability of this approach, we start with two simple models of one degree of freedom, the exactly solvable Gaussian case and a quartic model, both with a complex mass parameter. 
In each case we have generated training data by solving the discretised CL process with stepsize $\epsilon$ using the higher-order algorithm of Ref.~\cite{CCC:1987}, first applied to CL dynamics in Ref.~\cite{Aarts:2011zn}. This algorithm improves stepsize corrections from ${\cal O}(\eps)$ to ${\cal O}(\eps^{3/2})$. We have generated an ensemble of $10^6$ configurations for training, which are preprocessed by scaling it to zero mean and unit variance.

In the diffusion model, we employ a variance-exploding scheme, in which the  drift term in Eq.~(\ref{eq:forward_process}) is put to zero, $K(x,y;t) = 0$. Note that due to the complexification, we have two degrees of freedom to consider. We choose the diffusion coefficient $g(t) = \sigma^{t/T}$ and pick $\sigma = 10$ and $T=1$.
To model the score as $s_\theta(x,y;t)$, we use a time-conditioned fully connected neural network using Gaussian Fourier feature mapping \cite{tancik2020:GaussianFourier}. We choose to run the backward process using 1000 steps for $10^6$ trajectories to obtain samples. Our choice of hyperparameters is summarised in table \ref{tab:model_training_hyperparams}. More details can be found in Ref.~\cite{Aarts:2024rsl}.

\begin{table}[h!]
\centering
\begin{tabular}{ll|ll}
\toprule
\textbf{Hyperparameter}    & \textbf{Value}         & \textbf{Hyperparameter}    & \textbf{Value}         \\ \midrule
Layers                     & [64, 64]               & Learning Rate              & 1e-4                   \\
Time Embedding dims        & 128                    & Batch Size                 & 512                    \\
Activation Function        & LeakyReLU              & Optimizer                  & Adam                   \\
Weight Initialization      & LeCun Uniform \cite{LeCun1998}    & Max Epochs                 & 200                    \\ \bottomrule \\
\end{tabular}

\caption{Model and training hyperparameters used in training. We save the weights with the best loss during the training process and employ early stopping.}
\label{tab:model_training_hyperparams}
\end{table}

\subsection{Gaussian model}

As a first example, we consider the Gaussian action with complex mass parameter
\be
S(x) = \half\sigma_0 x^2, \qqquad \sigma_0 = A+iB.
\ee
The CL dynamics with real noise is described by the system of equations,
\be
\dot x = K_x +\eta, \qqquad  K_x=-Ax+By, \qqquad
\dot y = K_y, \qqquad K_y=-Ay-Bx.
\ee
The FPE admits a stationary solution \cite{Aarts:2009hn,Aarts:2015tyj}
\be
\label{eq:FPE-G}
P(x,y) = N\exp\left[ -\alpha x^2-\beta y^2-2\gamma xy\right],
\qqquad 
N = \frac{1}{\pi}\sqrt{\alpha\beta-\gamma^2},
\ee
with the coefficients 
$\alpha = A,
\beta = A(1+2A^2/B^2),
\gamma = A^2/B$.
This solution satisfies Eq.~(\ref{eq:rhoP}).
With this solution, the log derivatives of the probability distribution (\ref{eq:FPE-G}) sampled by the CL process can be computed and the analytical score reads
\be
\label{eq:dlogPexact}
\partial_x\log P(x,y)= -2\alpha x - 2\gamma y, \qqquad
\partial_y\log P(x,y)= -2\beta y - 2\gamma x.
\ee
As we see below, the learned score at the end of the backward process approximates this vector field. 
It is important to note that this distribution and its derivatives are usually not available and in particular not directly related to the CL drift $(K_x, K_y)$, 
since the latter cannot be integrated. This is easy to see, as $\partial_y K_x\neq \partial_xK_y$. 

\begin{figure}[t]
    \centering
    \includegraphics[width=\linewidth]{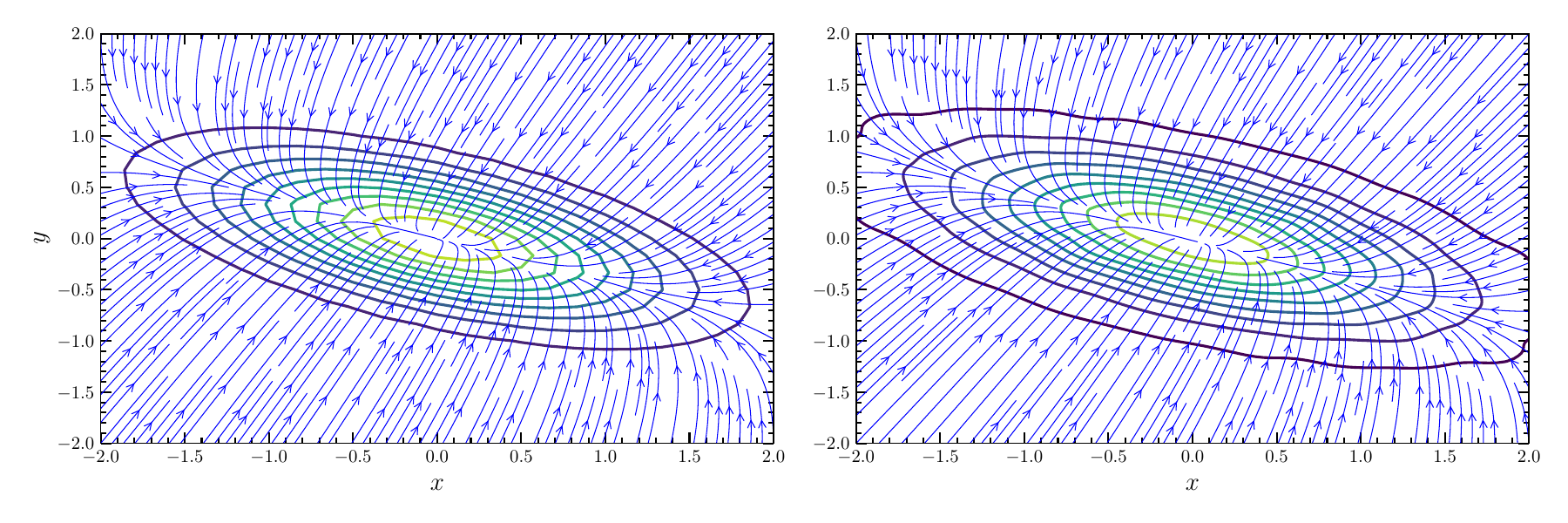}
    \caption{Gaussian model with complex mass parameter $\sigma_0 = 1+i$: 
    vector field $\nabla \log P(x,y)$ and lines of constant $P(x,y)$ as given by the analytical results (\ref{eq:FPE-G}, \ref{eq:dlogPexact}) (left) 
    and as learnt by the diffusion model (right).}
    \label{fig:Analytic_vs_DM_Gaussian}
\end{figure}

We have trained the diffusion model on data generated using CL dynamics, as explained above. The resulting score at the end of the backward process is shown in Fig.\ \ref{fig:Analytic_vs_DM_Gaussian} (right). This score can be compared with the analytical score (\ref{eq:dlogPexact}), shown on the left. Contour lines of constant $P(x,y)$ are included as well. We observe that the model manages to capture the score from the data. To make this more quantitative, we have computed the four lowest nonzero moments, $\mu_n = \E[(x+iy)^n]$ with $n=2,4,6,8$. The exact results are $\mu_n = (n-1)!!/\sigma_0^{n/2}$ for even $n$. Since the theory is Gaussian, all higher-order cumulants vanish. The results are shown in Table \ref{table:Gaussian}. Note that the diffusion model learns from CL generated data, not from the exact distribution.

\begin{table}[h!]
    \centering
    \scalebox{0.87}{
    \begin{tabular}{c|cc|cc|cc|ccc}
        \hline\hline
        $n$ & \multicolumn{2}{c}{2} & \multicolumn{2}{c}{4} & \multicolumn{2}{c}{6} & \multicolumn{2}{c}{8} \\
        \hline\hline
        & re & $-$im & re & $-$im & re & $-$im & re & $-$im \\
        Exact & 0.5 & 0.5 & 0 & 1.5 & $-$3.75 & 3.75 & $-$26.25 & 0 \\
        CL & 0.4986(7) & 0.4990(7) & $-$0.0018(1) & 1.494(5) & $-$3.75(2) & 3.75(3) & $-$26.4(3) & 0.20(3) \\
        DM & 0.497(1) & 0.491(1) & 0.021(1) & 1.476(7) & $-$3.65(3) & 3.78(4) & $-$26.3(1) & 0.81(68) \\   
         \hline\hline
    \end{tabular}
    }
    \caption{Gaussian model with complex mass parameter $\sigma_0 = 1+i$: first four non-vanishing moments $\mu_n$, as obtained from CL data and from diffusion model generated data, including exact values. Statistical errors are computed by a bootstrap resampling of the dataset with $10^6$ configurations using 100 bins.}
    \label{table:Gaussian}
\end{table}

\subsection{Quartic model}

We now consider the quartic model with a complex mass parameter \cite{Aarts:2013uza}
\be
 S = \half\sigma_0 x^2+\frac{1}{4}\lambda x^4, \qqquad\sigma_0 = A+iB.
\ee
Exact results can be obtained by a direct evaluation of the partition function,
\be
 Z = \int dx\, e^{-S(x)} = \sqrt{\frac{4\xi}{\sigma_0}}e^{\xi} K_{-\frac{1}{4}}(\xi),
\ee
 where $\xi=\sigma_0^2/(8\lambda)$ and $K_q(\xi)$ is the modified Bessel 
function of the second kind. Subsequently, moments $\mu_n = \E[x^n]$ are obtained by differentiating with respect to $\sigma_0$. Odd moments vanish.

Provided that $3A^2-B^2>0$, the CL process is contained in a strip $-y_-<y<y_-$, with \cite{Aarts:2013uza}
\be
y_-^2 = \frac{A}{2\lambda}\left( 1- \sqrt{1-\frac{B^2}{3A^2}}\right).
\ee
CL dynamics then yields the correct results \cite{Aarts:2013uza}.

\begin{figure}[t]
    \centering
    \includegraphics[width=.49\linewidth]{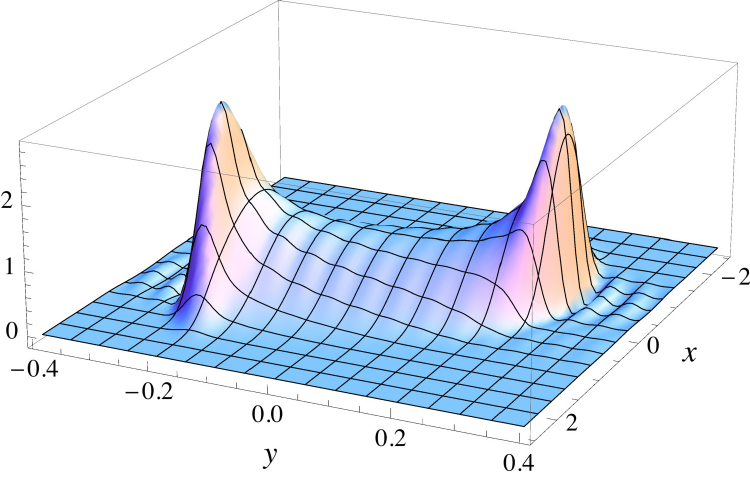}
    \includegraphics[width=.49\linewidth]{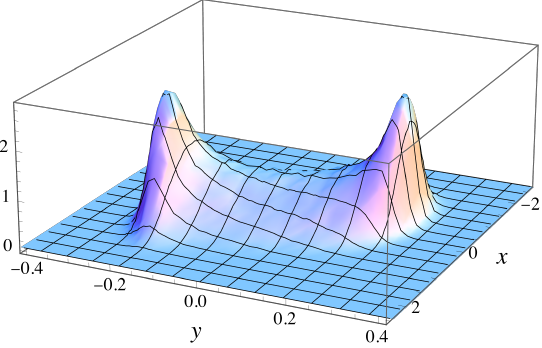}
    \caption{Quartic model with parameters $\sigma_0=1+i, \lambda=1$: solution $P(x,y)$ of the FPE obtained by a double expansion in Hermite functions \cite{Aarts:2013uza} (left) and as learnt by the diffusion model (right).}
    \label{fig:twopeaks}
    \centering
    \includegraphics[width=\linewidth]{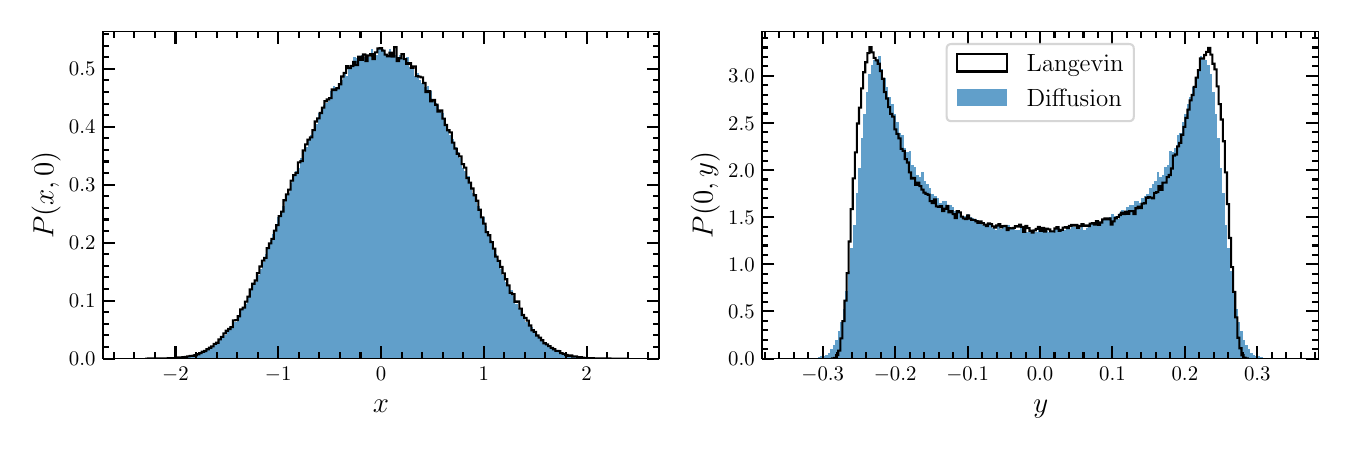}
    \caption{Distributions $P(x,0)$ and $P(0,y)$ created by collecting configurations during the CL process and from the trained diffusion model, using $10^6$ samples in each case. Other parameters as above. }
    \label{fig:Quartic_DM_Distributions}
\end{figure}

We have trained the diffusion model as above. In this case, no analytical expression for $P(x,y)$ or the score is available. In Ref.~\cite{Aarts:2013uza} the FPE was solved by a double expansion in terms of Hermite functions. The resulting stationary distribution is shown in Fig.\ \ref{fig:twopeaks} (left). The distribution is strictly zero when $|y|> y_- \simeq 0.3029$. The little ripples in the left plot are an artifact of the expansion. 
Sampling from the trained diffusion model yields the distribution shown on the right. 
Two cross sections of the distribution, $P(x,0)$ and $P(0,y)$, are shown in Fig.~\ref{fig:Quartic_DM_Distributions}. We note that the trained diffusion model manages to capture the two peaks characteristic of this model as well as the boundary restrictions from the training data, with some small deviations visible.

\begin{figure}[t]
    \centering
    \includegraphics[width=\linewidth]{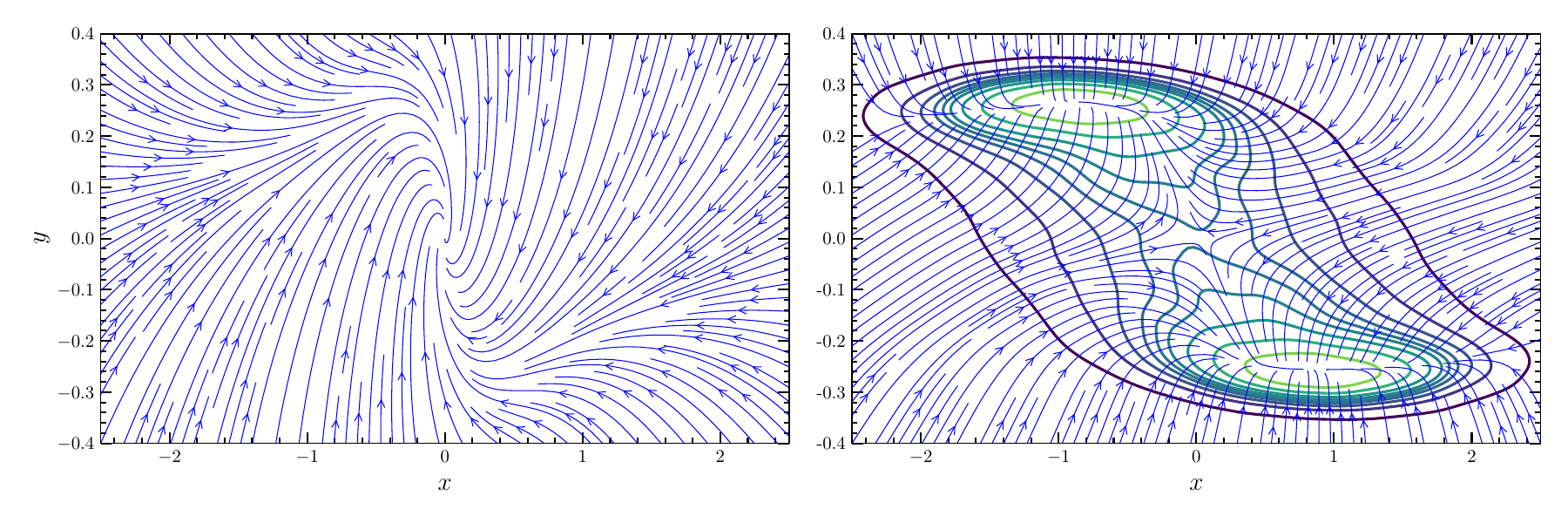}
    \caption{Drift in complex Langevin dynamics (left) and the score as learnt by the diffusion model, including contour lines (right). 
    These vector fields are different, as they should be. Parameters as above. }
    \label{fig:CL_vs_DM_quartic}
\end{figure}

The analytical score is not available in this case. In Fig.~\ref{fig:CL_vs_DM_quartic} we show the CL drift (left) and the score at the end of the backward process, as learnt by the diffusion model (right). The two vector fields are different, as they should be.
Recall that the Langevin drift is used in the CL equation with noise in the $x$ direction only and no time-dependent coefficients, whereas the score is used in the time-dependent stochastic equation with noise applied in both directions. 
Neverthless, both processes yield (approximately) the same distribution, used for data generation. 
To make the comparison quantitative, we have computed cumulants $\kappa_n$ from the numerically estimated moments $\mu_n$, using the standard relations.
The results are presented in Table \ref{tab:quartic_cumulants}. We observe good agreement.

\begin{table}[t]
    \centering
    \scalebox{0.87}{
    \begin{tabular}{c|cc|cc|cc|ccc}
        \hline\hline
        $n$ & \multicolumn{2}{c}{2} & \multicolumn{2}{c}{4} & \multicolumn{2}{c}{6} & \multicolumn{2}{c}{8} \\
        \hline\hline
        & re & $-$im & re & $-$im & re & $-$im & re & $-$im \\
        Exact   & 0.428142  & 0.148010  & $-$0.060347  & $-$0.100083  & $-$0.00934 & 0.19222 & 0.41578 & $-$0.5923 \\
        CL      & 0.4277(5) & 0.1478(2) & $-$0.0597(6) & $-$0.0991(6) & $-$0.010(1) & 0.188(2) & 0.406(4) & $-$0.57(1) \\
        DM      & 0.4267(6) & 0.1459(2) & $-$0.0582(6) & $-$0.0981(5) & $-$0.008(1) & 0.188(2) & 0.400(5) & $-$0.58(1) \\   
      \hline\hline
    \end{tabular}
    }
    \caption{As in Table \ref{table:Gaussian}, cumulants $\kappa_n$ for the quartic model with parameters $\sigma_0=1+i, \lambda=1$.}
    \label{tab:quartic_cumulants}
\end{table}

\section{Outlook}

We have demonstrated the diffusion model's ability to capture the distributions from data generated by complex Langevin dynamics, using two simple models. The model reproduces statistical properties of the data on which it is trained, as one would expect.
The capability of diffusion models to learn higher-order cumulants has also been shown in Ref.~\cite{Aarts:2024rsl}.
In the context of CL dynamics, a diffusion model will not solve the sign problem when CL fails. 
However, the score learnt by the diffusion model is not related to the drift in the CL process, but is instead an approximation to the gradient of an effective action on the complexified space. This may open up new avenues to analyse the properties of CL generated distributions on the complexified space, which are worth exploring. 
Finally, there are no obstacles to extend this approach to two-dimensional lattice field theories \cite{Wang:2023exq, Zhu:2024kiu} and use it to generate additional configurations.

\noindent
{\bf Acknowledgements} --  
DEH is supported by the UKRI AIMLAC CDT EP/S023992/1.
GA is supported by STFC Consolidated Grant ST/T000813/1. 
KZ is supported by the CUHK-Shenzhen University development fund under grant No.\ UDF01003041 and UDF03003041, and Shenzhen Peacock fund under No.\ 2023TC0179.

\noindent
{\bf Research Data and Code Access} --
The code and data used for this manuscript is a variation of the code available in Ref.~\cite{zenodo}.

\noindent
{\bf Open Access Statement} -- For the purpose of open access, the authors have applied a Creative Commons Attribution (CC BY) licence to any Author Accepted Manuscript version arising.

\providecommand{\href}[2]{#2}\begingroup\raggedright\endgroup

\end{document}